\documentclass[fleqn,10pt]{wlscirep}
\title{Optimal efficiency of the Q-cycle mechanism around physiological temperatures from an open quantum systems approach}

\usepackage{booktabs}
\usepackage{multirow}
\usepackage{siunitx}
\author[1,*]{Francesco Tacchino}
\author[2,3,$\dagger$]{Antonella Succurro}
\author[2,3]{Oliver Ebenh\"{o}h}
\author[1,+]{Dario Gerace}
\affil[1]{Department of Physics, University of Pavia, I-27100 Pavia, Italy}
\affil[2]{Institute of Quantitative and Theoretical Biology, Heinrich Heine University, 40225 D\"{u}sseldorf, Germany}
\affil[3]{Cluster of Excellence on Plant Sciences (CEPLAS), Heinrich Heine University, 40225 D\"{u}sseldorf, Germany}
\affil[*]{francesco.tacchino01@universitadipavia.it}
\affil[+]{dario.gerace@unipv.it}
\affil[$\dagger$]{Current address: Life and Medical Sciences (LIMES) Institute and  West German Genome Center, University of Bonn, Bonn, Germany}

\newcommand{\pq}{$\mathrm{PQ}$}
\newcommand{\pqh}{$\mathrm{PQH}_2$ }
\newcommand{\onlinecite}[1]{\hspace{-1 ex} \nocite{#1}\citenum{#1}}

\begin{abstract}
The Q-cycle mechanism entering the electron and proton transport chain in oxygenic photosynthesis is an example of how biological processes can be efficiently investigated with elementary microscopic models. 
Here we address the problem of energy transport across the cellular membrane from an open quantum system theoretical perspective. We model the cytochrome $b_6 f$ protein complex under cyclic electron flow conditions starting from a simplified kinetic model, which is hereby revisited in terms of a quantum master equation formulation and spin-boson Hamiltonian treatment. 
We apply this model to theoretically demonstrate an optimal thermodynamic efficiency of the Q-cycle around ambient and physiologically relevant temperature conditions. Furthermore, we determine the quantum yield of this complex biochemical process after setting the electrochemical potentials to values well established in the literature. The present work suggests that the theory of quantum open systems can successfully push forward our theoretical understanding of complex biological systems working close to the quantum/classical boundary.
\end{abstract}
\begin{document}

\flushbottom
\maketitle

\thispagestyle{empty}

\section*{Introduction}

Approaching the field of biological sciences from the perspective of microscopic physical processes is extremely intriguing and yet partly unexplored in current research. 
Presently, photosynthesis occupies a particularly privileged position in this quest: the molecular basis of its fundamental mechanisms are actively investigated at the biological, biochemical, and biophysical levels. Additionally, the nature of the process itself is well suited to be investigated with the tools of either classical or quantum statistical physics. In fact, it involves analyzing the light-matter interaction within the cell as well as excitation and particle transport through molecular networks, together with compelling energetic considerations. It has been more than ten years now since the attention of quantum physicists was triggered by remarkable spectroscopic results reporting possible evidence for quantum coherent processes in light harvesting complexes\ \cite{Engel2007}. We have henceforth witnessed the rapid development of the field of quantum biology\ \cite{Lambert2012,QuantumBioBook2014}, a growth that continues nowadays towards many directions at the forefront of pure and applied scientific research\ \cite{Creatore2013,Potocnik2017}. 

\begin{figure}
\centering
\includegraphics[scale=0.58]{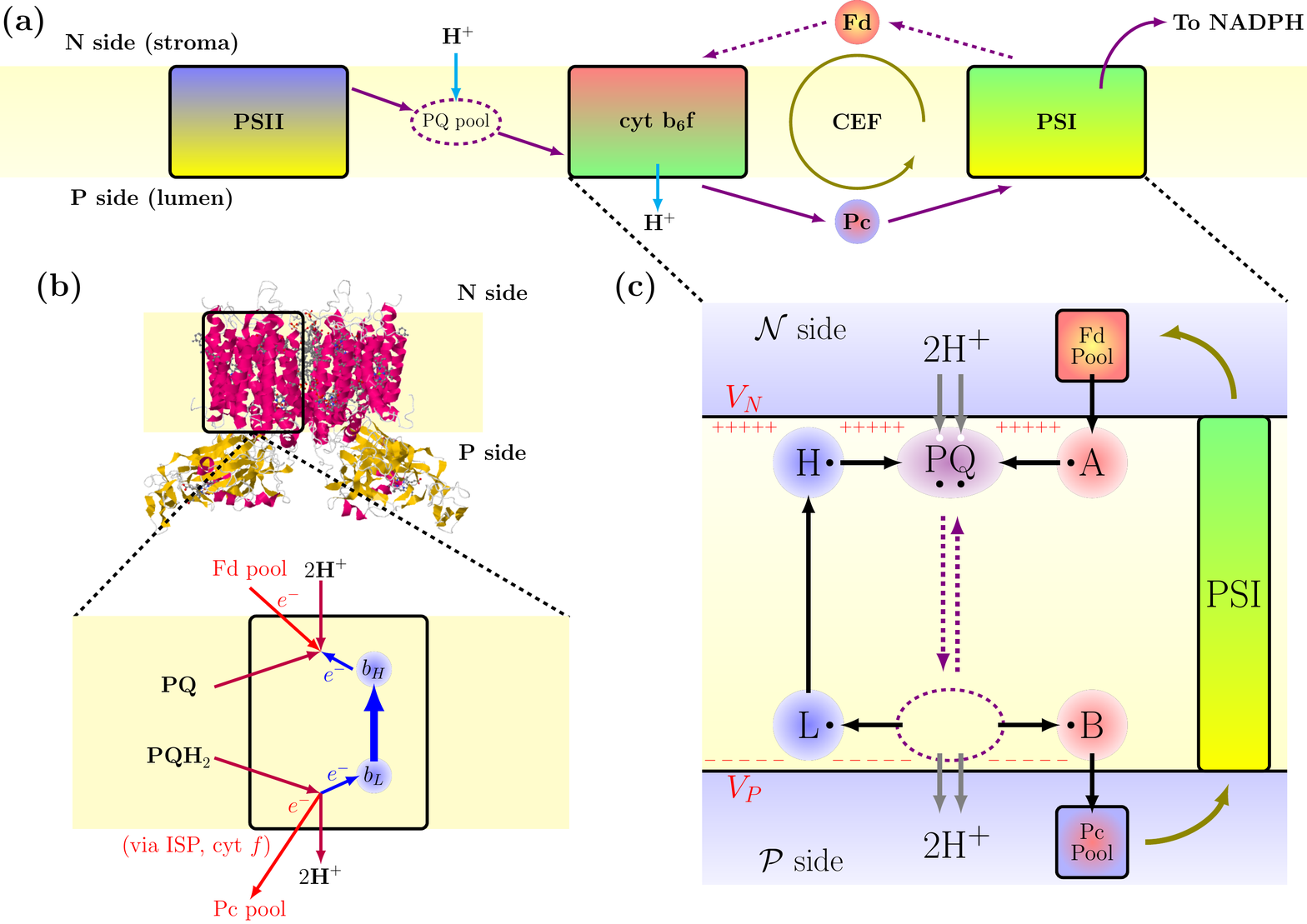} 
\caption{\textbf{Model for electron transport chain and cytochrome $b_6f$ complex}.
(a) The electron transport chain of oxygenic photosynthesis. Solid purple arrows show the ideal linear electron flow (LEF) in which electrons go through the whole chain and are finally transferred to a biological redox agent called NADPH. Dashed purple arrows show the alternative path that makes cyclic electron flow (CEF) possible. (b) Molecular structure of the cytochrome $b_6f$ complex (protein data bank, accession code PDB-1VF5) with a description of the Q-cycle mechanism under CEF conditions. The electron recycling chain is marked with blue arrows, and biological counterparts for the elements considered in the model are reported. (c) Pictorial representation of the model described in this article. Single electron binding sites are represented as small black circles, and proton ones as small white circles. The PQ purple element represents the plastoquinone/plastoquinol molecule, which is able to commute between N- and P-side reaction sites, as indicated by purple arrows. The black arrows show the ideal path of the electrons in the system.
}
\label{Fig:ETCandModel}
\end{figure}

While the significance and interpretation of many initial findings in quantum biology, most notably the origin and the effect of quantum coherences, is still actively debated \cite{Nalbach2011,Tiwari2013,Halpin2014,Duan2017,Palecek2017}, one of the most relevant theoretical achievements in the field lies in the application of the theory of open quantum systems to model biological processes featuring a single or a few excitations embedded in the complex landscape represented by a protein, a membrane, or even a cell \cite{QuantumBioBook2014}. Following a similar approach, we hereby focus on the electron transport chain, which represents the second stage of the photosynthetic machinery, immediately connected to the primary light-absorbing processes. Building on a kinetic model originally proposed by Smirnov and Nori\ \cite{Smirnov2012}, we develop a simple but accurate open quantum systems theory of the Q-cycle. The latter is a remarkable biochemical mechanism that efficiently conserves energy stored in excited electrons to increase the number of protons per electron that are pumped against the concentration gradient across the cellular membrane.

In deriving the relevant rate equations describing the dynamics of this complex biochemical process, we exploit the open quantum system formalism to describe single-particle transfer reactions on a microscopic level. At difference with previous derivations, our approach allows to consistently connect charge transport within the Q-cycle to the reservoir dynamics of the surrounding environment. 
After benchmarking our model by calculating the so-called quantum yield (i.e., the number of transferred protons per electron during each cycle) and comparing our results to similar ones already present in the literature\ \cite{Smirnov2012}, we then proceed to determine the thermodynamic efficiency of the Q-cycle as a function of the external temperature. We find optimal efficiency around ambient temperatures, a result that is remarkably compatible with physiological conditions for living cells. We further strengthen our analysis by introducing a new figure of merit combining information from the quantum yield and the transferred charge, which again shows optimality around room temperatures.
Finally, it should be emphasized that even though no genuine quantum effect such as entanglement or coherence-enhanced transport is expected to play any relevant role in the Q-cycle itself, the standard tools of quantum statistical mechanics prove to be well suited for the physical study of microscopic biochemical transport processes. 
Complemented by alternative mathematical approaches\ \cite{Ransac2008,Matuszynska2016}, this theoretical work potentially provides a deeper and more complete understanding of many biological aspects at the molecular level.

\section*{Model and theoretical description}
\label{sec:model}

In Fig.{~}\ref{Fig:ETCandModel}a we show a concise diagram of the full photosynthetic electron transport chain. Light is absorbed and stored in the form of chemical energy inside large proteins called photosystem I (PSI) and photosystem II (PSII), respectively~\cite{Blankenship2014}. These photosystems are typically assisted by light harvesting complexes (LHC) during the photon absorption and exciton migration stage. 
The two photosystems are actually successive stages of the same global energy transfer chain in which the energy contained in high energy chemical bonds fuels the transfer of protons ($\mathrm{H}^+$) against their concentration gradient across the membrane of the photosynthetic cellular organelles, the ``thylakoids''. This leads to a non-equilibrium distribution of protons, where the proton concentrations on the P-side (``lumen'') is orders of magnitude larger than on the N-side (``stroma''). 
Such out-of-equilibrium unbalance results mainly from two processes: first, the electrons radiatively excited in PSII are extracted from water, leading to the release of oxygen and protons in the lumen; second, the chemical energy of excited electrons is used to transfer protons against their concentration gradient, a process orchestrated by a third membrane-spanning molecular complex, the so-called cytochrome $b_6f$ complex, which will be the object of our analysis in the following of this work.

The cytochrome $b_6f$ complex works essentially as an electron-fueled proton pump implementing the Q-cycle mechanism, as originally proposed by P.\ Mitchell~\cite{Mitchell1975,Mitchell1975a} and successively discussed and adapted~\cite{Crofts2004,Crofts2008,Mulkidjanian2010,Cramer2011}. 
A simplified kinetic model of the Q-cycle mechanism was originally introduced in Ref.~\onlinecite{Smirnov2012}. The full details of such implementation are reported in the Methods section, for completeness. We now present our original formulation, which is built on the same elementary ingredients but makes in addition explicit use of the language and methods of open quantum systems theory. Indeed, by introducing the evolution of the total density matrix of the system, we can describe consistently the microscopic origin of all the contributions to the dynamics, including the interaction with external reservoirs. The model describes the proton and electron motions through different binding sites, which are individually treated as two-level systems. The proposed use of quantum theory allows for a straightforward derivation of the time evolution equations,  directly arising from the definition of all the possible elementary states of the system and their mutual interactions. At the same time, thanks to the intrinsically open system character of our formulation, the interaction between the microscopic details of single-particle transfer reactions can be very naturally and operatively connected to the mesoscopic and reservoir dynamics of the surrounding degrees of freedom.  We make use of the pseudospin\ $s = 1/2$ formulation, in which the occupied Fock state $|1\rangle$ indicates the presence of a particle on the specific site. Throughout the paper, quantum mechanical operators for electron binding sites will be written with small letters (e.g.\ annihilation operators $a_i$), while we will use capital letters for protons (e.g.\ annihilation operators $A_i$). Annihilation operators obey anticommutation rules $\{a_i,a_j^\dagger\} = \delta_{ij}$ and $\{A_i,A_j^\dagger\} = \delta_{ij}$. For simplicity of notation, we will assume $\hbar = k_B = 1$ in the following.

In Fig.{~}\ref{Fig:ETCandModel}b-c we provide a schematic description of the working principles of the cytochrome $b_6f$ complex under the so called cyclic electron flow (CEF) conditions, a situation that turns out to be simpler to model and to simulate, while keeping all the interesting features of the Q-cycle mechanism. Several detailed structural descriptions of the cytochrome $b_6f$ complex exist in the literature~\cite{Kurisu2003,Stroebel2003,Cramer2006,Cramer2006a,Baniulis2008}, including all the large heme and other prosthetic groups embedded inside the protein scaffold acting as electron binding sites. In particular, it is currently accepted that  two reaction sites are present at the two opposite sides of this membrane-spanning protein complex, respectively called $Q_o$ towards the interior of the thylakoid (lumen) and $Q_i$ on the other side (stroma), where specific mobile electron carriers, called plastoquinone/plastoquinol ($\mathrm{PQ}$/$\mathrm{PQH}_2$) can fit one per site at a time. Each $\mathrm{PQ}$ molecule (purple in Fig.{~}\ref{Fig:ETCandModel}c) can be either discharged or charged with up to two electrons (mainly coming from PSII) and two protons, thus transforming into a $\mathrm{PQH}_2$. In principle, a whole pool of \pq/\pqh molecules is present inside the membrane, to which these hydrophobic species are confined while being free to diffuse, but only few of them can be present at the same time inside the body of the $b_6f$ complex. For simplicity, a single plastoquinone/plastoquinol molecule is included in our model, described as a carrier of up to two protons and two electrons, which can spatially commute between the N- and P-side of the cellular membrane and alternatively interacting with the $Q_i$ an $Q_o$ reaction sites. We will often refer to this component as the PQ-shuttle. The Hamiltonian describing the microscopic degrees of freedom of such plastoquinone contains multiple contributions, namely $H_Q = H_e + H_p + H_{ee} + H_{pp} + H_{ep}$ (details on the individual terms can be found in the Methods section): the first two terms describe two pairs of independent binding sites for electrons and protons, respectively, while the other terms describe the mutual electrostatic interactions between the charged particles simultaneously present on the shuttle. These are intended as an effective way to take into account the differences in free energy (\textit{i.e.}\ in standard redox potential) between different molecular species such as quinone, semiquinone, and quinol.

When a \pqh enters the $Q_o$ reaction site, it interacts with two electron binding sites, namely heme-$b_L$ and the Iron-Sulfur (ISP) domain ($B$ in Fig.{~}\ref{Fig:ETCandModel}c). One electron leaves the \pqh and is transferred through the ISP via cytochrome $f$ to a Plastocyanin (Pc) molecule, a water-soluble single electron carrier acting as a connector with PSI. Hence, the original redox energy of such electron is completely consumed. One proton is released to the lumen side (also called P-side for proton rich), leaving the original plastoquinol in a semiquinone state. The second electron is transferred to heme $b_L$ and then to heme $b_H$ across the membrane, while the second proton is released to the P-side: the plastoquinol is now in the fully oxidized plastoquinone state. On the other side of the membrane, such plastoquinone molecules can bind to the $Q_i$ site close to heme $b_H$. The electron that traversed the $L$-$H$ chain reduces this plastoquinone to a semiquinone state. A second fresh electron is provided directly by PSI through a water soluble single-electron carrier called Ferredoxin (Fd) which can bind to the N-side of the $b_6f$ complex. Two protons are taken up from the stromal side, resulting in a fully reduced $\mathrm{PQH}_2$, ready to diffuse back into the membrane and towards the P-side. The hemes $b_L$ and $b_H$ that constitute an electron-recycling chain are again described by a composite Hamiltonian containing free energy terms, $H_{LH, free}$, accompanied by a Coulomb repulsive interaction, $H_{LH, int}$, which reflects the redox anticooperativity of the two hemes, as measured in molecular complexes structurally and functionally similar to the $b_6f${~}\cite{Kim2012}. Referring to Fig.~\ref{Fig:ETCandModel}c, the $B$ site collectively represents the ISP-cytochrome $f$ chain that transfers electrons to the Pc pool. On the other hand, site $A$ models the electron re-injection site from the Fd pool to \pq. Each of the $A$ and $B$ sites can bind a single electron, while the Ferredoxin and Plastocyanin pools on the stromal and lumenal sides, respectively, are modeled as collections of fermionic oscillators. We assume that PSI transfers electrons from the Plastocyanin pool to the Ferredoxin pool with the help of external energy (i.e., from light absorption), and that this mechanism is at a steady state, such that pools can be described with time-independent parameters. The protons in the bulk aqueous phase on the P- and N-side can be treated in a similar way. With a circuital analogy, the Fd and Pc pools act as source and drain leads for the proton pump, while the \pq/\pqh molecules are used as mobile parts. It is easy to see that, as a result of the set of reactions described above, the overall process translocates two protons to the P-side per electron passed to the Pc pool. Therefore, under ideal conditions, the quantum yield (QY) of the reaction, defined as the ratio between the number of protons released on the P-side and the number of electrons whose redox energy is consumed approaches the theoretical value 
\begin{equation}
\mathrm{QY} _\mathrm{ideal} = 2.
\end{equation}
This is particularly relevant in view of calculating the thermodynamic efficiency of the proton translocation process, defined as the ratio between energetic outputs and inputs, which is proportional to QY~\cite{Smirnov2012}:
\begin{equation}
\eta = \frac{\Delta \mu_{\mathrm{prot}}}{\Delta\mu_\mathrm{el}}\mathrm{QY}
\label{eq:eff_general}
\end{equation} 
where $\Delta\mu$ denotes the change in electrochemical potential for electrons and protons.

We now solve for the dynamics of this model by explicitly resorting to the theory of open quantum systems applied to the full density matrix of the complete system, $\rho$. Notice that the total dimension of the Hilbert space corresponds to a collection of 8 independent two-level systems, i.e. $d = 2^8$. We hereby describe $A$ and $B$ sites interacting with the electron source and drain in terms of typical Lindblad terms (with $(i,j)\in \{(Fd,A),(Pc,B)\}$)~\cite{Breuer2002}
\begin{equation}
\mathcal{L}_{i} [\rho] = \bar{n}_{ij} \frac{\gamma_{i}}{2} \Theta (a^\dagger_j)[\rho] + (1-\bar{n}_{ij}) \frac{\gamma_{i}}{2} \Theta (a_j)[\rho]
\label{eq:FdPcLindblad}
\end{equation}
where the Lindblad dissipator is $\Theta \left(V\right)[\rho] = 2V\rho V^\dagger - \left\{V^\dagger V, \rho\right\}$, and the average occupation numbers of the reservoirs are described by Fermi-Dirac distributions at the $A$ and $B$ energies $\epsilon_j$
\begin{equation}
\bar{n}_{ij} = f(\epsilon_j; T, \mu_{i}) = \left[ \exp\left(\frac{\epsilon_j-\mu_{i}}{T}\right) +1 \right]^{-1}
\end{equation}
where $T$ is the temperature and $\mu_i$ are the source and drain electrochemical potentials, respectively. 
The electron transfer reactions can be phenomenologically modeled through incoherent Lindblad-type terms including forward and backward Marcus rates \cite{Marcus1956,Marcus1985}. A complication arises here, since the single Marcus rates from two individual states explicitly depend on their energy difference. Therefore, we shall distinguish between the cases in which other charged molecules are present or not, for example in all the transitions involving shuttle electrons.
This is precisely the effect of the Coulomb interaction terms, which contribute to make some specific transitions more or less favorable. We thus introduce projectors on the states of the system, $P_i = |i\rangle\langle i |$, $i = 1,..., 2^8$, and add the following contributions to the master equation
\begin{equation}
\mathcal{L}_{i\rightarrow j}^{x\rightarrow y} [\rho] = \frac{1}{2} \chi_{xy}(i,j) \Theta (P_j a_x a_y^\dagger P_i)[\rho]
\label{eq:lindblad_marcus}
\end{equation}
to describe the transition from state $i$ to state $j$ via the tunneling connection from site $x$ to site $y$. For simplicity, we assume that the reorganization energies only depend on the sites and not on the states of the system: we can therefore write
\begin{equation}
\chi_{xy} (i,j) = |\Delta_{xy}|^2 \sqrt{\frac{\pi}{\lambda_{xy} T}} \exp \left(\frac{-(\lambda_{xy}+ \omega_j - \omega_i)^2}{4\lambda_{xy} T}\right)
\end{equation}
where $\omega_i$ is the energy eigenvalue of state $|i\rangle$ as an eigenstate of the free Hamiltonian of the system, and $\Delta_{xy}$ is a tunneling matrix element. Finally, a similar problem also occurs in defining the interaction between the proton sites on the PQ-shuttle and the N- and P-side reservoirs. In this case, it is the Fermi-Dirac distribution describing the average occupation number that depends on the energy difference of each specific transition. The solution comes again with the help of state projectors:
\begin{equation}
\mathcal{L}_{i\rightarrow j}^{\alpha,k} [\rho] = f(\Omega_{ij}; T, \mu_{\alpha}) \frac{\Gamma_\alpha}{2} \Theta (P_j A^\dagger_k P_i)[\rho] + (1-f(\Omega_{ij}; T, \mu_{\alpha})) \frac{\Gamma_\alpha}{2} \Theta (P_i A_k P_j)[\rho]
\end{equation}
where $\alpha = \text{N,P}$, $k=1,2$ and $\Omega_{ij}$ is the absolute value of the energy difference between the states $i$ and $j$. If we now define
\begin{equation}
\mathcal{L}_{Prot}^{\alpha,k} [\rho] = \sum_{ij} \mathcal{L}_{i\rightarrow j}^{\alpha,k} [\rho]
\qquad \mathcal{L}_{Elec}^{x\rightarrow y} [\rho] = \sum_{ij} \mathcal{L}_{i\rightarrow j}^{x\rightarrow y} [\rho] 
\end{equation}
we can write the full evolution of the system in the form of a single master equation
\begin{equation}
\frac{d\rho}{dt} = \mathcal{L}_{Fd} [\rho] + \mathcal{L}_{Pc} [\rho]+ \sum_{x \neq y} \mathcal{L}_{Elec}^{x\rightarrow y} [\rho] + \sum_{\alpha,k} \mathcal{L}_{Prot}^{\alpha,k} [\rho]
\label{eq:model_master}
\end{equation}
where  we impose $\chi_{xy}(i,j) = 0\,\,\forall i,j$ if sites $x$ and $y$ are not directly connected in the model.

As already anticipated, from Eq.{~}\eqref{eq:model_master} it clearly appears that the off-diagonal entries of the density matrix evolve independently from the populations. The effect of quantum coherence is thus dynamically ineffective here, and it can safely be neglected when considering the electron transfer dynamics. Since we are dealing with a fully incoherent picture, we can then recast the relevant part of Eq.{~}\eqref{eq:model_master} in the form of a Pauli master equation \cite{Breuer2002}: indeed, the dynamics of the diagonal elements of the density matrix, namely the electron and proton populations in the possible basis states, evolve according to a relaxation equation of the form
\begin{equation}
\frac{d\mathbf{P}}{dt} = \Lambda_{Relax} \mathbf{P}
\label{eq:matrix_relax}
\end{equation}
where $\mathbf{P}$ is a vector with $2^8$ components describing the populations (diagonal elements of $\rho$), and the time-independent relaxation matrix, $\Lambda_{Relax}$, receives contributions from all the components of Eq.{~}\eqref{eq:model_master}. Notice that, at difference with the original approach~\cite{Smirnov2012}, the present formalism is based on the complete density matrix of the system and retains the linear structure typical of quantum mechanics, even at the stage where it is reduced to a set of rate equations. Therefore, it fully captures all possible correlation effects (in principle, both at the quantum and semiclassical level) without further assumptions. In particular, it conserves the total number of particles in electron and proton transfer reactions. Moreover, since the full master equation gives in principle access to the dynamics of the quantum mechanical coherences, the formalism could be easily extended to include a richer phenomenology when applied to different contexts.

A very peculiar element of the description of the $b_6f$ complex under CEF conditions is the stochastic modeling of the PQ shuttle motion inside the membrane between the N- and the P-side, as already introduced~\cite{Smirnov2012}. 
Here we keep such an approach, thus including a stochastic differential equation (SDE) to describe the position of the PQ shuttle, whose diffusive commuting inside the rather dense lipidic membrane is modeled as an overdamped Langevin-Brownian motion
\begin{equation}
\zeta \dot{x} = -\frac{dU_w}{dx} - \langle(n_1 + n_2 - N_1 - N_2)^2\rangle \frac{dU_{ch}}{dx} + \xi
\label{eq:stoch_motion}
\end{equation}
where $\zeta$ is the drag coefficient (\textit{i.e.}\ the inverse of the diffusion coefficient times the temperature, $\mathcal{D}=T/\zeta$) and $\xi$ represents a gaussian noise source with zero average and correlation function $\langle\xi(t)\xi(t')\rangle = 2\zeta T \delta(t-t')$. The potential $U_w$ is added in order to confine the shuttle inside the membrane, while $U_{ch}$ prevents an electrically charged shuttle from crossing the lipid core of the membrane. The tunneling rates between the electron sites on the shuttle and the sites $A$, $B$, $L$, and $H$ will then depend on the position of the shuttle. This can be described with an exponential decay of the coupling rate~\cite{Blankenship2014}
\begin{equation}
\Delta_{iQ} (x) = \Delta_{iQ} \exp \left(\frac{x\pm x_0}{l_e}\right)
\end{equation}
with the lower sign for $i = \text{L, B}$ and the upper sign for $i = \text{H, A}$. Here $x_0 = 2\,$nm, and $l_e$ is a characteristic length. Since the shuttle can only interact with the aqueous phase of the N- or P-side when it is spatially close to the membrane border, we assume position-dependent rates
\begin{equation}
\Gamma_i (x) = \Gamma_i \exp \left(\frac{x \pm x_0}{l_p}\right)
\end{equation}
where the upper sign refers to $i=\text{N}$ and the lower to $i=\text{P}$, respectively, thus distinguishing the two sides of a membrane with a total thickness of $4\,$nm. Finally, the distribution of charged aminoacid residues is at the origin of a structural asymmetry in the electrical surface potential of the $b_6f$ complex, which is taken into account by adding an internal potential between the membrane boundaries $\pm x_0$. Such a potential is linearly varying between two temperature-dependent extreme values $V_N=4.6\,$T and $V_P=5.4\,$T,  and its effect is to rescale the free excitation energies of the electron and proton binding sites (see the Methods section), as already described~\cite{Smirnov2012}.

\begin{figure*}[t]
\centering
\includegraphics[scale=0.26]{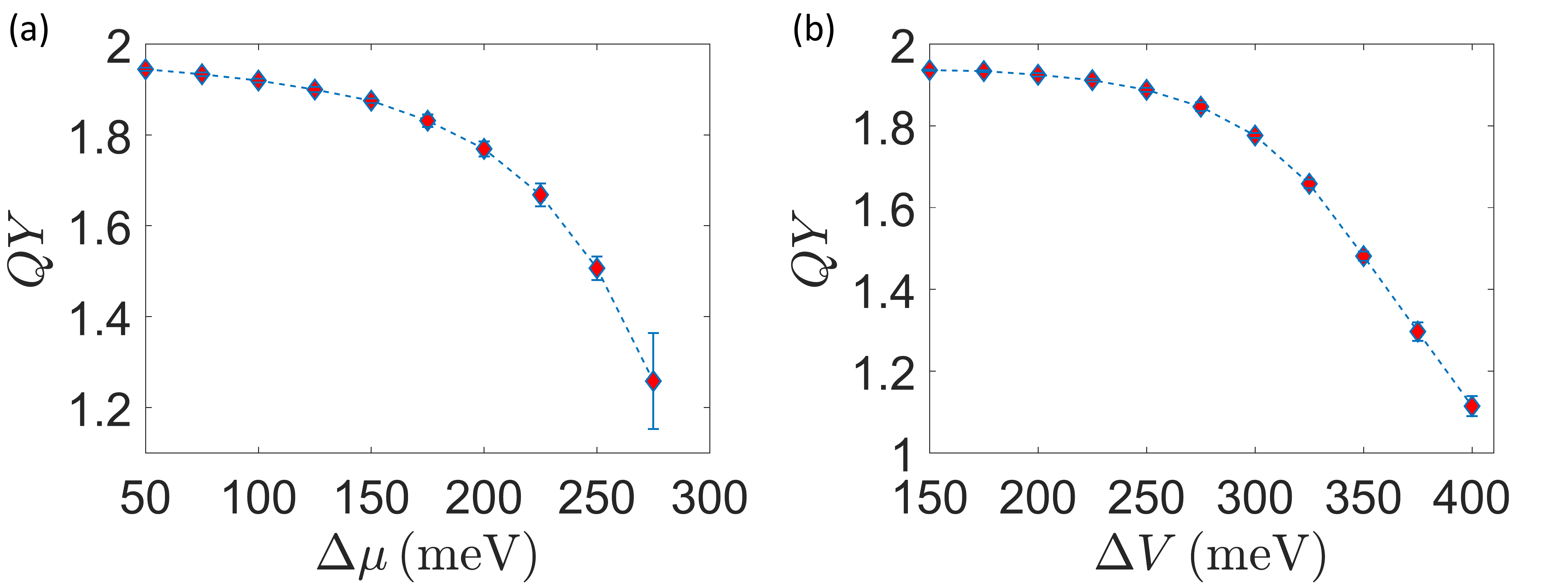}
\caption{\textbf{Benchmark simulations}.
(a) Quantum Yield as a function of the electrochemical gradient of protons (b) Quantum Yield as a function of the electrostatic surface potential.}
\label{Fig:Benchmark}
\end{figure*}

\begin{figure*}[t]
\centering
\includegraphics[scale=0.26]{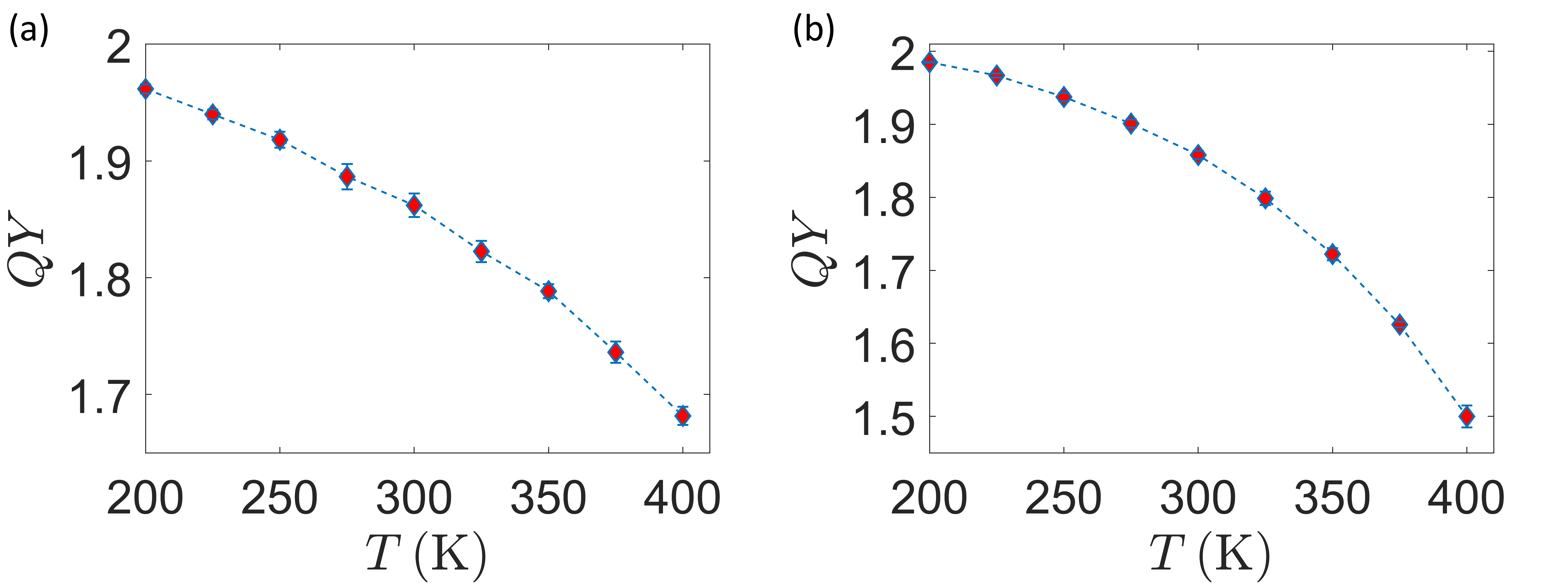}
\caption{\textbf{Quantum Yield as a function of temperature}.
(a) Quantum Yield as a function of $T$ for fixed $\Delta\mu$ of protons and $\Delta V$ (scheme I) (b) Quantum Yield as a function of $T$ for fixed $\Delta$pH of protons and $\Delta V$ (scheme II).}
\label{Fig:T1}
\end{figure*}

\section*{Results}

\begin{figure*}[t]
\centering
\includegraphics[scale=0.23]{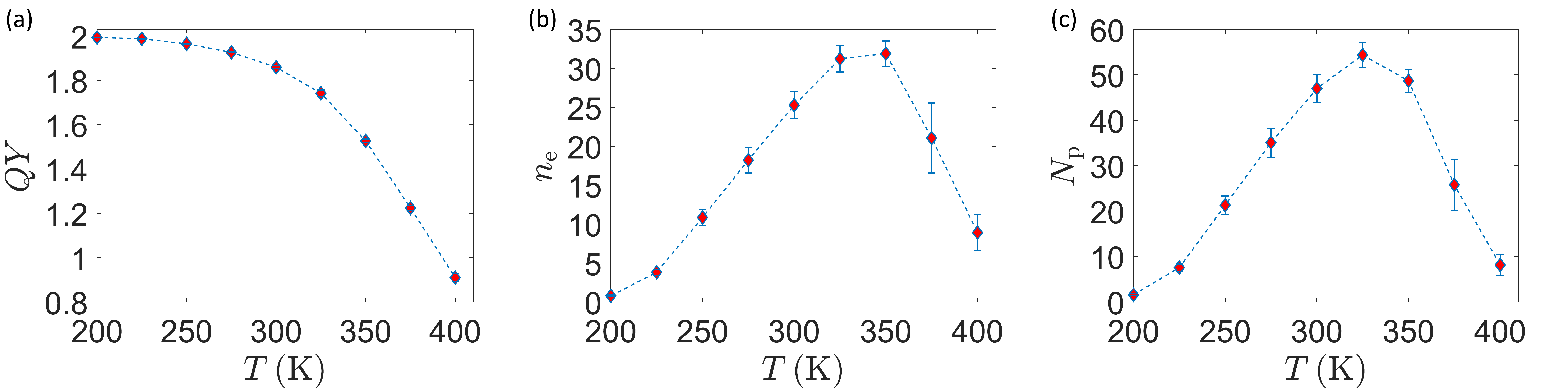}
\caption{\textbf{Temperature scan at constant $\Delta$pH and varying surface potential}.
(a) Quantum Yield as a function of $T$ for fixed $\Delta$pH of protons and with $\Delta V$ varying with temperature (scheme III) (b)-(c) Average total number of translocated electrons ($n_e$) and protons ($N_p$) during a full trajectory under scheme III.}
\label{Fig:T2}
\end{figure*}

\begin{figure*}[t]
\centering
\includegraphics[scale=0.26]{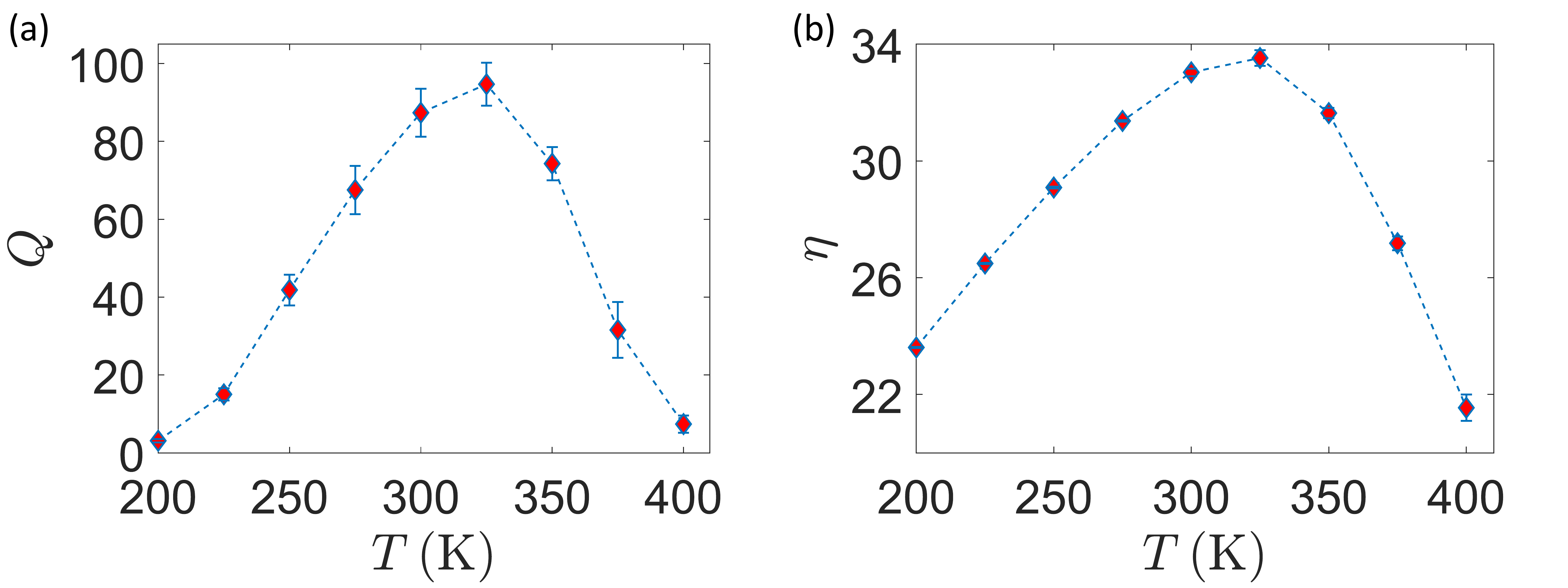}
\caption{\textbf{Figures of merit under scheme III}.
(a) Quantum Yield times the number of translocated protons (b) Thermodynamic efficiency.}
\label{Fig:Qeff}
\end{figure*}

In this section, we present some of the most significant results we obtained by performing extensive numerical simulations of the hybrid quantum and stochastic model discussed above (see also details in the Methods section). Notice that all the simulations presented in this work can be performed on a standard Personal Computer.
The dynamical behavior of the system was investigated by varying some physically meaningful external parameter, such as the electrochemical gradient of protons across the membrane, the static surface potential, or the temperature. For each parameter under individual scan, the relevant range of variation was empirically identified and divided into a number of steps. For each one of these steps, i.e. for each individual value of the parameter under scan, six independent stochastic diffusion trajectories were simulated, all with the same set of parameters and initial conditions. Each trajectory covered $100\,\mu$s of $b_6f$ complex activity. The mean value and standard deviation of the quantum yield and the total number of transferred electrons and protons (i.e.\ the translocation rate) were then computed for each sampling point from the outputs of the six stochastic realizations.

A few benchmark simulations were performed in order to validate our formalism with respect to previous results~\cite{Smirnov2012}. In all the simulations, any parameter not involved in the scan was kept fixed at a default value chosen consistently with the relevant literature~\cite{Smirnov2012}, in such a way to guarantee a consistent comparison of the results for benchmarking our theoretical method. All such values are reported in Table \ref{tab:parameters}, see the Methods section for completeness. First, the electrochemical gradient difference for protons, $\Delta \mu = \mu_P - \mu_N$, was gradually varied while symmetrically keeping $\mu_N = - \mu_P$. Since the following relation generally holds~\cite{NF2002}
\begin{equation}
\Delta\mu \approx -\Delta \mathrm{pH} \frac{T}{298\,\text{K}} (60\,\text{meV})
\label{eq:deltamu_pH}
\end{equation}
at a fixed temperature the scan over $\Delta\mu$ is equivalent to a scan over $\Delta$pH, namely the concentration gradient of $\mathrm{H^+}$. The results for the simulated QY are shown in Fig.{~}\ref{Fig:Benchmark}a. As it can easily be seen, the system is stable and keeps good performances over a wide range of values, with QY well above 1 and close to the ideal value, $\mathrm{QY} = 2$. The quantum yield decreases when the concentration gradient against which the protons are pumped becomes very large. Remarkably, these results show good agreement with previously reported ones~\cite{Smirnov2012}. 
As a second benchmark simulation, the electrostatic surface potential difference $\Delta V = V_P + V_N$ was varied at constant $\Delta\mu$ and $T$. At room temperature ($T_R = 298\,$K) we have $V_P = 140\,\text{meV}$ and $V_N = 120\,\text{meV}$ with $V_P - V_N = 20\,\text{meV}$. Following Smirnov and Nori, we varied $\Delta V$ while keeping $V_P - V_N$ constant. The results are presented in Fig.{~}\ref{Fig:Benchmark}b. Again, there is good agreement with the results previously reported~\cite{Smirnov2012}.
Here we go beyond previous studies by explicitly checking for the effects of temperature variations. Temperature is explicitly included in the definition of many quantities affecting the dynamical behavior of the model, and influences both electron and proton transfer reactions as well as the stochastic motion of the $\mathrm{PQ}$ shuttle and other structural parameters.  In our simulations, the temperature was scanned according to three different schemes, summarized in the following. 

In scheme I, the temperature was varied while keeping fixed $\Delta\mu$ for protons and the surface potential. In such a case, temperature variations only affect the reservoir populations given by Fermi-Dirac distributions and the Marcus electron-transfer rates. Moreover, the diffusion coefficient for the stochastic motion of the $\mathrm{PQ}$ shuttle is also affected. According to Eq.~\eqref{eq:deltamu_pH}, the pH gradient also depends on $T$ at constant $\Delta\mu$. Within this formulation, we analyzed the response of the system mainly for what intrinsically concerns the electron and proton transfer dynamics, all the other conditions and properties being unchanged. The results for this scan are presented in Fig.{~}\ref{Fig:T1}a.

In scheme II, we held the pH gradient fixed while changing the temperature. This time, the $\Delta\mu$ for protons changes at different sampling points, again according to Eq.~\eqref{eq:deltamu_pH}. This situation  is of great biological interest, since realizes the case in which the pH is under external control. At each point, given the temperature and $\Delta\mathrm{pH} = 2.5$, $\Delta\mu$ is obtained from Eq.~\eqref{eq:deltamu_pH} and then distributed as $\mu_P = \Delta\mu / 2$ and $\mu_N = -\Delta\mu / 2$. The results of this scan, as shown in Fig.{~}\ref{Fig:T1}b, are not radically different from the previous ones in Fig.{~}\ref{Fig:T1}a: we can thus infer that the impact of a temperature change on the particle dynamics inside the system dominates over other factors, such as the $\Delta\mu$ change. 
As the third and more complete stage (scheme III) we also took into account the temperature dependence of the surface electrostatic potential. As already shown~\cite{Smirnov2012}, this dependence takes the form $V_P = 5.4 T$ and $V_N = 4.6 T$. Again, we kept $\Delta$pH fixed and plotted the results in Fig.{~}\ref{Fig:T2}. The connection between $\Delta\mu$ and $\Delta$pH was not modified, assuming that ion balance across the thylakoid membrane always provides a net zero electrostatic component to the protonic electrochemical potential.
The $T$-scan simulations show that the quantum yield decreases while increasing the temperature, while the number of translocated protons increases. If achieving a quantum yield as high as possible is certainly beneficial, this comes at the cost of a reduced number of transferred protons per unit time: in the biological optimization perspective, a trade-off should then be found at the point where the $b_6f$ activity is both sufficiently quick and efficient. A possible figure of merit is the product
\begin{equation}
Q = \text{QY}\cdot (\text{Number of transferred protons})
\end{equation}
which we expect to show a maximum in the intermediate range of temperatures. This is indeed the case, as shown in Fig.~\ref{Fig:Qeff}a for the data of simulation scheme III. Quite remarkably, the maximum of $Q$ is found in a temperature range that broadly lies around physiological conditions.

Finally, we can apply the formal general definition for thermodynamic efficiency, as given in Eq.~\eqref{eq:eff_general}, to the CEF case if we recognize that the input energy is provided by the electrochemical potential gradient of electrons from source (Fd pool) to drain (Pc pool), and that work is performed to move protons from the N- to the P-side. We thus have
\begin{equation}
\eta = \frac{\mu_P - \mu_N}{\mu_S - \mu_D} \text{QY}
\end{equation}
It is particularly interesting to compute such quantity when both the quantum yield and the electron-to-proton energy conversion ratio vary simultaneously. This is precisely the situation, for example, of the $T$ scan at constant pH and $T$-dependent surface potential, whose efficiency is plotted in Fig.~\ref{Fig:Qeff}b. Also in such a case where a purely energetic figure of merit is considered, we remarkably find a peak of the efficiency around the physiologically relevant temperatures of living organisms.

\section*{Discussion}

The Q-cycle mechanism is a relevant energy conversion process entering the electron transport chain in oxygenic photosynthesis, but its complex biochemical nature, difficult to capture with theoretical approaches, has to date limited the development of analytic and numerical models. Building on a microscopic description based on an open quantum system formalism, here we have started to bridge the gap between a first principles dynamical evolution of elementary charges and physiologically relevant conclusions on a macroscopic level. 
First, we were able to reproduce one of the most interesting effects associated with the Q-cycle, namely the bifurcation of electron flows: upon net transfer of two protons from one side (stroma) to the other (lumen) of the cellular membrane, only one of the two electrons simultaneously present on the charge carrier gets consumed, while the other goes through a recycling path within the $b_6f$ complex ready to trigger another cycle. 
Moreover, as an original result of this work we were able to capture a few biologically relevant conclusions regarding the efficiency of the mechanism itself and the physiological optimality of its design. 
In particular, we found that few biochemically relevant figures of merit show maximal values around ambient temperatures for typical values of the electrochemical potentials.

In general terms, it is remarkable that the Q-cycle mechanism is highly conserved, since not only it is found in the thylakoid membranes of photosynthetic organisms, but also in the inner mitochondrial matrix, where it serves as an energy-conserving proton-pumping mechanism in the $bc_1$ complex as part of the electron transport chain of oxidative phosphorylation. This points not only at the common evolutionary origin of these complexes and the Q-cycle mechanism, but also indicates that the energy conservation that they implement is highly beneficial to a wide variety of organisms. Our results further suggest that such systems might also be examined under the light of selective environmental adaptation.

In perspective, the present model is intended as a first step towards a full description of the $b_6f$ complex, and may serve as a starting point to investigate either the detailed pathway of CEF or its behavior in response to specific external conditions. 
Indeed, while it has been shown that the redox chemistry alone is sufficient to justify the electron bifurcation~\cite{Ransac2008}, some phenomenological observations, mainly obtained while studying the response of the mitochondrial $bc_1$ complex to the introduction of specific inhibitors, pointed towards the necessity of explicitly formulating some gating mechanisms~\cite{Rich2004,Osyczka2005,Ransac2010}. In principle, these hypotheses could be theoretically tested by extending the present model to include an external control over the allowed transitions, and better detailing the individual chemical mechanisms of such processes. 

On the technical side, we highlight that our work clearly demonstrates that an open quantum system formulation is sufficient to explain the Q-cycle as a naturally emerging process, given the general structural and physical parameters defining the protein complex environment. Moreover, such formalism is certainly suited for the exploration of more genuinely quantum effects at the mesoscale. Indeed, it is worth reminding that quantum mechanical features are naturally embedded into the description at a fundamental level: this is reflected in the linear structure of the resulting set of rate equations, which in principle preserve all possible quantum and semiclassical correlation effects and could readily be extended to situations in which the dynamics of quantum coherences plays a non-trivial role.

Finally, we cannot neglect that the interplay between biology and its formal analysis is bidirectional: in this respect, the elementary  CEF formulation adopted here might inspire new possible routes to artificially engineer coupled electron-proton translocations. 
 
\section*{Methods}
\label{app:methods}

Here we provide further details on the mathematical structure of the model described in the main text, together with the techniques that were used to derive and solve the resulting set of equations.

\textit{Structure of the microscopic model}. Here we explicitly report all the Hamiltonian terms entering the microscopic model of the Q-cycle discussed in the main text and further details concerning the stochastic motion of the $\mathrm{PQ}$ shuttle. As previously stated, these structural elements are chosen consistently with the original kinetic model in Ref.\ \onlinecite{Smirnov2012} and constitute the basis for our original developments based on an open quantum systems formalism. The explicit expression of the full system Hamiltonian can be written as
\begin{equation}
H = H_Q + H_{LH} + H_A + H_B + H_{Fd} + H_{Pc}
\end{equation}
For the $\mathrm{PQ}$ shuttle, we have $H_Q = H_e + H_p + H_{ee} + H_{pp} + H_{ep}$
where the electron and proton Hamiltonians are (here the indeces $i=1,2$ refer to the two different binding sites for electrons, with operators $a_i$, and protons, operators $A_i$)
\begin{equation}
\begin{split}
H_e = & \, \epsilon_Q \, a_1^\dagger a_1 + \epsilon_Q \, a_2^\dagger a_2 \\
H_p = & \, E_Q \, A_1^\dagger A_1 + E_Q \, A_2^\dagger A_2
\end{split}
\end{equation}  
while other terms describe the electrostatic interaction between the charged particles on the shuttle. Each site interacts independently with the others and the interaction is diagonal on the Fock basis, thus affecting only the energies of the states without inducing any transition:
\begin{equation}
\begin{split}
H_{ee} = & \, U_{ee} \, n_1n_2 \\
H_{pp} = & \, U_{pp} \, N_1N_2 \\
H_{ep} = & \, -U_{ep} \, (n_1 + n_2)(N_1 + N_2)
\end{split}
\end{equation}
where $n_i = a_i^\dagger a_i$ and $N_i = A_i^\dagger A_i$. The Hamiltonian $H_{LH}$ for the $L$ and $H$ sites also contains free energy terms and a Coulomb repulsive interaction
\begin{equation}
\begin{split}
H_{LH, free} = & \, \epsilon_L \, a_L^\dagger a_L + \epsilon_H \, a_H^\dagger a_H \\
H_{LH, int} = & \, U_{LH} \, n_L n_H
\end{split}
\end{equation}
Finally, $A$ and $B$ are treated as single-electron binding sites:
\begin{equation}
H_A = \epsilon_A \, a^\dagger_A a_A \qquad H_B = \epsilon_B \, a^\dagger_B a_B
\end{equation}
The Ferredoxin and Plastocyanin pool on the stromal and lumenal side respectively are modeled as collections of fermionic oscillators
\begin{equation}
\begin{split}
H_{Fd} = & \, \sum_k \epsilon_{k, Fd} \, c_{k, Fd}^\dagger c_{k, Fd} 
\\
H_{Pc} = & \, \sum_k \epsilon_{k, Pc} \, c_{k, Pc}^\dagger c_{k, Pc}
\end{split}
\end{equation}
Equivalent expressions can be given for the the P- and N-side proton reservoirs.

\begin{table*}[t]
\centering
\begin{tabular}{cScScS}
\hline
\multicolumn{4}{c}{\textbf{Energies and rates}} & \multicolumn{2}{c}{\textbf{Lengths and diffusion}} \\
\textbf{Parameter} & \textbf{Value (meV)} & \textbf{Parameter} & \textbf{Value (meV)} & \textbf{Parameter} & \textbf{Value (nm)}\\ 
\hline
$\epsilon_{Q0}$ & 280 & $T$ & 25 & $l_p $ & 0.25 \\
$E_{Q0}$ &  822 & $\mu_{Fd}$ &  410 & $l_e$ & 0.25 \\
$\mu_{N}$ &  -75 & $\mu_{P}$  &  75 & $x_{ch}$ & 1.70 \\
$V_P$ &  140 & $\mu_{Pc}$  &  -440 & $x_{w}$ & 2.70\\
$V_N$ &  120 & $\gamma_{Fd},\gamma_{Pc}$ & 0.0001 & $l_{ch}$ & 0.05\\
$\epsilon'_L$ & 360 & $\Gamma_P $ & 0.002 & $l_w $ & 0.10\\
$\epsilon'_H$ & 220 & $\Gamma_N$ & 0.002 &\textbf{Parameter} & \textbf{Value (meV$\,\mu$s$/$nm$^2$)}\\ \cline{5-6}
$\epsilon'_A$ & 465 & $\Delta_{AQ} $ & 0.10 & $\zeta$ & 8.55 \\
$\epsilon'_B $ & -495 & $\Delta_{BQ} $ & 0.10  & & \\
$\lambda_{AQ} $ & 100 & $\Delta_{LQ} $ & 0.06 & &\\
$\lambda_{BQ} $ & 100 & $\Delta_{HQ} $ & 0.06 & &\\
$\lambda_{LQ} $ & 100 & $\Delta_{LH} $ & 0.10 & &\\
$\lambda_{HQ} $ & 100 & $U_{ee} $ & 305 & &\\
$\lambda_{LH} $ & 250 & $U_{ep} $ & 610 & &\\
$U_{LH} $ & 240 & $U_{ch0} $ & 770 & &\\
$U_{pp} $ & 76.30 & $U_{w0} $ & 500 & &\\
\hline
\end{tabular}
\caption{Default parameters for the numerical simulations performed in this work.}
\label{tab:parameters}
\end{table*}

The asymmetric electrostatic surface potential, varying linearly with position inside the membrane, can be modeled as
\begin{equation}
V(x) = -\frac{x-x_0}{2x_0} V_N - \frac{x+x_0}{2x_0}V_P
\end{equation} 
The effect of $V(x)$ is to rescale the free excitation energies of the electron and proton binding sites. For sites $A$, $B$, $L$ and $H$ this is just a static shift
\begin{equation}
\epsilon'_{A/H} = \epsilon_{A/H} - V_N  \qquad \epsilon'_{B/L} = \epsilon_{B/L} + V_P \\
\end{equation}
while for the PQ-shuttle this means that the free energies become position-dependent
\begin{equation}
\epsilon_Q (x) = \epsilon_{Q0} - V (x) \qquad E_Q (x) = E_{Q0} + V (x)
\end{equation}
We assume that the surface potential does not affect the Brownian motion: indeed, it could only act on electrically non-neutral states of the shuttle and, in such case, the motion across the membrane is strongly suppressed by the action of the hydrophobic energy barrier $U_{ch}$. In other words, the stochastic random force originating from molecular collision is dominant in governing the diffusion  of the $\mathrm{PQ}$ shuttle. The explicit analytic form of the confining potentials for the diffusion are the following:
\begin{equation}
\begin{split}
U_w (x) = & \, U_{w0}\left\{1-\left[\exp\left(\frac{x-x_w}{l_w}\right)+1\right]^{-1}+\left[\exp\left(\frac{x+x_w}{l_w}\right)+1\right]^{-1}\right\} \\
U_w (x) = & \, U_{ch0}\left\{\left[\exp\left(\frac{x-x_{ch}}{l_{ch}}\right)+1\right]^{-1}-\left[\exp\left(\frac{x+x_{ch}}{l_{ch}}\right)+1\right]^{-1}\right\}
\end{split}
\end{equation}

\begin{figure*}[t]
\centering
\includegraphics[scale=0.25]{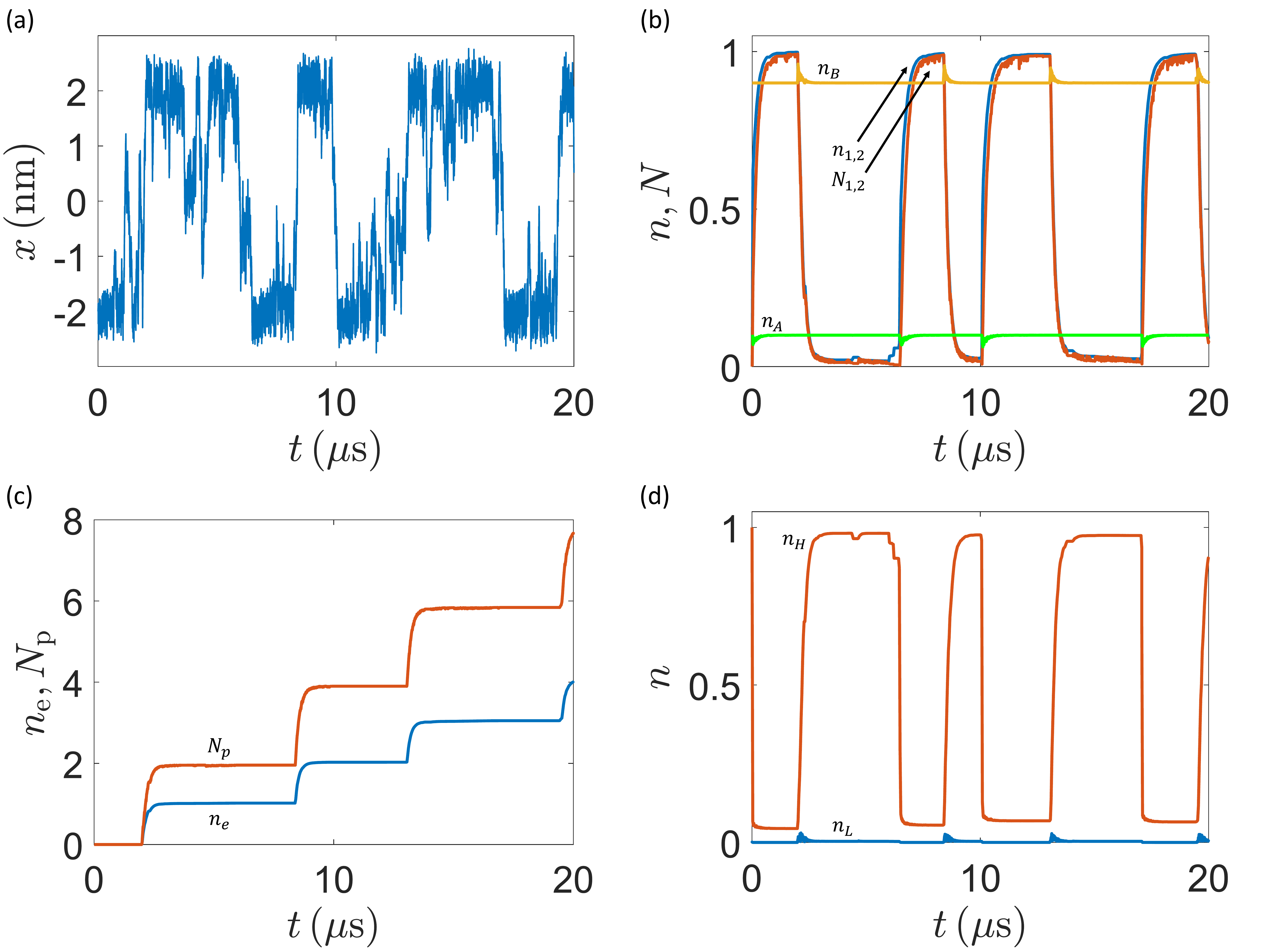}
\caption{\textbf{Typical stochastic trajectory of the model system}.
(a) Position of the $\mathrm{PQ}$ shuttle inside the membrane. (b)-(d) Electron ($n_i$) and proton ($N_i$) average population on the $\mathrm{PQ}$ shuttle and on the $A$, $B$, $L$ and $H$ sites. Notice the spikes corresponding to the charging and discharging of the shuttle when it approaches the N- and P-side, as well as the remarkable coupling between electron and proton motion (the latter are translocated against their electrochemical gradient). (c) Total number of electrons transferred to the Pc pool ($n_e$) and total number of protons ($N_p$) brought from the N- to the P-side.}
\label{Fig:traj}
\end{figure*}

\textit{Open quantum systems treatment}. The master equation for the quantum evolution was derived by using techniques from the theory of open quantum systems. The source (Fd) and drain (Pc) terms are realizations of the well known case of a single two-level system in contact with a thermal reservoir whose creation and annihilation operators obey anti-commutation rules. The proton reservoir terms are similar in spirit but distinguish the possible configurations of electrons and protons on the $\mathrm{PQ}$ shuttle, essentially recognizing that the difference in redox potential is well resolved despite the line-broadening induced by the interaction with the reservoirs. Finally, the Marcus transition rates can be derived starting from Hamiltonian terms of the form
\begin{equation}
H = \epsilon_k a_k^\dagger a_k + \epsilon_l a_l^\dagger a_l + \Delta (a_k^\dagger a_l + h.c.) + \sum_j \frac{p_j^2}{2m_j} + \frac{m_j \omega_j^2}{2} [x_j - x_{j,k}n_k - x_{j,l}n_l]^2
\end{equation}
which are essentially an adapted version of the well known spin-boson Hamiltonian{~}\cite{Leggett1987} featuring the two binding sites coupled by a tunneling interaction and connected to the surrounding molecular vibrational environment. Marcus transition rates are obtained by applying perturbation theory and Fermi's golden rule while assuming a Debye form for the bath spectral function $J(\omega) \propto \omega/(1 + \omega^2\tau^2)$ (see Ref.{~}\onlinecite{Xu1994} for details).

\textit{Numerical solution}. The dynamics of the system was numerically simulated with an original Python code.
The coupled quantum and stochastic system equations
\begin{equation}
\begin{cases}
\frac{d\mathbf{P}}{dt} = \, & \Lambda_{Relax}(x) \mathbf{P} \\
\zeta \dot{x} = \, & -\frac{dU_w}{dx} - \langle(n_1 + n_2 - N_1 - N_2)^2\rangle \frac{dU_{ch}}{dx} + \xi
\end{cases}
\end{equation}
display two separate dominant time-scales, namely the fast (on the order of picoseconds) quantum dynamics and the slower (on the order of fractions of microseconds) mesoscopic diffusion of the $\mathrm{PQ}$ shuttle. At every time step $dt = 10^{-3}\,\mu$s, we computed the local energy values and we evolved the populations from the previous state. Notice that, at difference with the formally similar set of rate equations reported in Ref. \onlinecite{Smirnov2012}, here the fully linear structure of Eq.{~}\eqref{eq:matrix_relax} makes it possible to use matrix exponentiation to compute the exact time evolution for arbitrary times
\begin{equation}
\mathbf{P}(t+t_0) = e^{\Lambda_{Relax}t} \mathbf{P}(t_0)
\end{equation}
We used the above expression at each step for a time $dt$, and we updated all the relevant observables, including the net charge on the shuttle, making use of the standard quantum mechanical formalism for the expectation values. At the end of such step, the position was updated using an Euler scheme for the integration of the SDE. The default parameters that we used in all the simulations, unless otherwise specified in the text, are summarized in Tab.{~}\ref{tab:parameters}. All of them are mostly consistent with the values already known from the relevant literature~\cite{Smirnov2012}.

As an explicit example, we show in Fig.~\ref{Fig:traj} part of a typical stochastic trajectory of the system dynamical evolution, as obtained with the default parameters. 
The corresponding time evolution of the electron and proton populations is also reported, for completeness.

\section*{Acknowledgements}

This project was partly supported from the EU Commission through the Erasmus+ program, the Cluster of Excellence on Plant Sciences (CEPLAS), and the Institute of Advanced Study (IUSS) of Pavia.
A.S.\ and O.E.\ acknowledge funding from the European Commission Seventh Framework Marie Curie Initial Training Network project AccliPhot (grant agreement PITN-GA-2012-316427). This study was funded by the Deutsche Forschungsgemeinschaft (DFG) under Germany’s Excellence Strategy EXC 2048/1, Project ID: 390686111.

%
%
%
%

\end{document}